\documentclass[onecolumn,11pt]{article}
\usepackage[top=1in, bottom=1in, left=1in, right=1in]{geometry}
\usepackage{amsmath,amsfonts,amscd,amssymb}
\usepackage{graphicx}
\usepackage{subfigure}
\usepackage{epstopdf}
\usepackage{overpic}
\usepackage{arcs}
\usepackage{color}
\usepackage{setspace}
\usepackage{palatino} 
\usepackage{relsize}
\usepackage{lmodern}
\usepackage{slantsc}
\usepackage{float}
\usepackage{hyperref}
\usepackage{comment}
\usepackage{xcolor}
\usepackage{color,soul}

\usepackage[numbers,sort&compress]{natbib}

\definecolor{offlinered}{RGB}{115, 0, 0}
\definecolor{onlineblue}{RGB}{4, 79, 149}

\DeclareSymbolFont{yhlargesymbols}{OMX}{yhex}{m}{n}
\DeclareMathAccent{\wideparen}{\mathord}{yhlargesymbols}{"F3}

\usepackage{gensymb}

\usepackage{amsmath}
\usepackage{tikz,calc}
\makeatletter
\newlength\@SizeOfCirc%
\newcommand{\CricArrowRight}[1]{%
\setlength{\@SizeOfCirc}{\maxof{\widthof{#1}}{\heightof{#1}}}%
\tikz [x=1.0ex,y=1.0ex,line width=.15ex, draw=black]%
\draw [->,anchor=center]%
node (0,0) {#1}%
(0,0.8\@SizeOfCirc) arc (90:-90:0.8\@SizeOfCirc);%
}%
\makeatother

\usepackage{amsfonts} 
\DeclareMathSymbol{\shortminus}{\mathbin}{AMSa}{"39}

\newenvironment{mat}{\left[\begin{array}{ccccccccccccccc}}{\end{array}\right]}
\newcommand\bcm{\begin{mat}}
\newcommand\ecm{\end{mat}}

\newenvironment{rmat}{\left[\begin{array}{rrrrrrrrrrrrr}}{\end{array}\right]}
\newcommand\brm{\begin{rmat}}
\newcommand\erm{\end{rmat}}

\newcommand{\eql}{\begin{equation}\label}

\UseRawInputEncoding

\begin{document}
\title{\vspace{-.1in}\huge{Learning physics-constrained subgrid-scale closures in the small-data regime for stable and accurate LES}}

\author{\normalsize{Yifei Guan$^1$\thanks{yifei.guan@rice.edu}, Adam Subel$^1$, Ashesh Chattopadhyay$^1$, and Pedram Hassanzadeh$^{1,2}$\thanks{pedram@rice.edu}}\\
\footnotesize{$^1$Department of Mechanical Engineering, Rice University, Houston, TX, 77005, United States}\\
\footnotesize{$^2$Department of Earth, Environmental and Planetary Sciences, Rice University, Houston, TX, 77005, United States}\\
}
\date{}
\maketitle

\maketitle

\begin{abstract}
We demonstrate how incorporating physics constraints into convolutional neural networks (CNNs) enables learning subgrid-scale (SGS) closures for stable and accurate large-eddy simulations (LES) in the small-data regime (i.e., when the availability of high-quality training data is limited). Using several setups of forced 2D turbulence as the testbeds, we examine the {\it a priori} and {\it a posteriori} performance of three methods for incorporating physics: 1) data augmentation (DA), 2) CNN with group convolutions (GCNN), and 3) loss functions that enforce a global enstrophy-transfer conservation (EnsCon). While the data-driven closures from physics-agnostic CNNs trained in the big-data regime are accurate and stable, and outperform dynamic Smagorinsky (DSMAG) closures, their performance substantially deteriorate when these CNNs are trained with 40x fewer samples (the small-data regime). An example based on a vortex dipole demonstrates that the physics-agnostic CNN cannot account for never-seen-before samples$'$ rotational equivariance (symmetry), an important property of the SGS term. This shows a major shortcoming of the physics-agnostic CNN in the small-data regime. We show that CNN with DA and GCNN address this issue and each produce accurate and stable data-driven closures in the small-data regime. Despite its simplicity, DA, which adds appropriately rotated samples to the training set, performs as well or in some cases even better than GCNN, which uses a sophisticated equivariance-preserving architecture. EnsCon, which combines structural modeling with aspect of functional modeling, also produces accurate and stable closures in the small-data regime. Overall, GCNN+EnCon, which combines these two physics constraints, shows the best {\it a posteriori} performance in this regime. These results illustrate the power of physics-constrained learning in the small-data regime for accurate and stable LES.
\end{abstract}

\section{Introduction}
Large-eddy simulation (LES) is widely used in the modeling of turbulent flows in natural and engineering systems as it offers a balance between accuracy and computational cost. In LES, the large-scale structures are explicitly resolved on a coarse-resolution grid while the subgrid-scale (SGS) eddies are parameterized in terms of the resolved flow using a closure model~\cite{smagorinsky1963general,pope2001turbulent,sagaut2006large,moser2020statistical}. Therefore, the fidelity of LES substantially depends on the accuracy of the SGS closure, improving which has been a longstanding goal across various disciplines~\cite[e.g.,][]{bou2005scale,sagaut2013multiscale,schneider2017earth, moser2020statistical,zanna2021deep}. In general, the SGS models can be classified into two categories: structural and functional~\cite{sagaut2006large}. Structural models, such as the gradient model~\cite{leonard1975energy,clark1979evaluation}, are developed to capture the structure (pattern and amplitude) of the SGS stress tensor and are known to produce high correlation coefficients $(c)$ between the true and predicted SGS terms ($c>0.9$) in \emph{a priori} analysis. However, they often lead to numerical instabilities in \emph{a posteriori} LES, for example, because of excessive backscattering and/or lack of sufficient dissipation~\cite{meneveau2000scale,lu2013modulated,vollant2016dynamic,wang2021dynamic,yuan2021dynamic}. Functional models, such as the Smagorinsky model~\cite{smagorinsky1963general} and its dynamic variants~\cite{germano1991dynamic,lilly1992proposed,zang1993dynamic}, are often developed by considering the inter-scale interactions (e.g., energy transfers). While producing low $c$ ($< 0.6$) between the true and predicted SGS terms in \emph{a priori} analysis~\cite{pawar2020priori,guan2021stable,wang2018investigations,zhou2019subgrid}, these functional models usually provide numerically stable \emph{a posteriori} LES, at least partly due to their dissipative nature. Thus, developing SGS models that perform well in both \emph{a priori} and \emph{a posteriori} analyses has remained a long-lasting research focus. \\

In recent years, there has been a rapidly growing interest in using machine learning (ML) methods to learn data-driven SGS closure models from filtered direct numerical simulation (DNS) data~\cite[e.g.,][]{wang2017physics,duraisamy2019turbulence,maulik2019sub,beck2019deep,pawar2020priori,prat2020priori,taghizadeh2020turbulence,zanna2020data,brunton2020machine,beck2021perspective,duraisamy2021perspectives,moriya2021inserting,portwood2021interpreting,stoffer2021development,liu2021investigation,jiang2021interpretable,tian2021physics}. Different approaches applied to a variety of canonical fluid systems have been investigated in these studies. For example, Maulik~\textit{et al.}~\cite{maulik2019subgrid,maulik2019sub} and Xie~\textit{et al.}~\cite{xie2019modeling,xie2019artificial,xie2019artificial2} have, respectively, developed {\it local} data-driven closures for 2D decaying homogeneous isotropic turbulence (2D-DHIT) and 3D incompressible and compressible turbulence using multilayer perceptron artificial neural networks (ANNs); also see \cite{zhou2019subgrid,xie2021artificial,wang2021artificial,maulik2021turbulent,sonnewald2021bridging}. Zanna and Bolton~\cite{bolton2019applications,zanna2020data}, Beck and colleagues~\cite{beck2019deep,kurz2020machine}, Pawar~\textit{et al.}~\cite{pawar2020priori}, Guan~\textit{et al.}~\cite{guan2021stable}, and Subel~\textit{et al.}~\cite{subel2021data} developed {\it non-local} closures, e.g., using convolutional neural networks (CNNs), for ocean circulation, 3D-DHIT, 2D-DHIT, and forced 1D Burgers' turbulence, respectively. While finding outstanding results in \emph{a priori} analyses, in many cases, these studies also reported instabilities in \emph{a posteriori} analyses, requiring further modifications to the learnt closures for stabilization. \\

More recently, Guan~\textit{et al.}~\cite{guan2021stable} showed that increasing the size of the training set {\it alone} can lead to stable and accurate \emph{a posteriori} LES (as well as high $c$ in \emph{a priori} analysis) even with {physics-agnostic} CNNs\footnote{For simplicity, we refer to any ``physics-agnostic CNN'' as {CNN} hereafter.}. This was attributed to the following: Big training datasets obtained from filtered DNS (FDNS) snapshots can provide sufficient information such that the physical constraints and processes (e.g., backscattering) are correctly learnt by data-driven methods, leading to a stable and accurate LES in both \emph{a priori} and \emph{a posteriori} analyses. However, in the {\it small-data} regime, the physical constraints and processes may not be captured correctly, and the inaccuracy of the data-drivenly predicted SGS term (particularly inaccuracies in backscattering) can result in unstable or unphysical LES~\cite{guan2021stable}. Note that as discussed later in Section~\ref{sec:method}, whether a training dataset is {\it big} or {\it small} depends on both the number of samples and the inter-sample correlations; thus it depends on the total length of the available DNS dataset, which can be limited due to computational constraints. As a result, there is a need to be able to learn data-driven closures in the small-data regime for stable and accurate LES. This can be achieved by incorporating physics into the learning process, which is the subject of this study. \\

Past studies have shown that embedding physical insights or constraints can enhance the performance of data-driven models, e.g., in reduced-order models~\cite[e.g.,][]{achatz1997closure,loiseau2018constrained,wan2018data,guan2021sparse,kaptanoglu2021physics,vinuesa2021potential,khodkar2021data,maulik2022aieada} and in neural networks~\cite[e.g.,][]{wu2018physics,king2018deep,sharma2018weakly,pan2018long,chattopadhyay2020analog,meidani2021data,beucler2021enforcing,charalampopoulos2021machine,subel2021data,prakash2021invariant,tian2021physics,yan2021data,magar2021auglichem,pawar2022frame}. There are various ways to incorporate physics in neural networks (e.g., see the reviews by \citet{kashinath2021physics}, \citet{balaji2021climbing}, and \citet{karniadakis2021physics}). For neural network-based data-driven SGS closures, in general, three of the main ways to do this are: data augmentation (DA), physics-constrained loss functions, and physics-aware network architectures. Training datasets can be constructed to represent some aspects of physics. For example, Galilean invariance and some of the translational and rotational equivariances of the SGS term~\cite{pope2001turbulent,silvis2017physical} can be incorporated through DA, i.e., built into the input and output training samples~\cite{prat2020priori,frezat2021physical,subel2021data}. Here, ``equivariance" means that the SGS terms are preserved under some coordinate transformations, resulting from properties of the Navier-Stokes equations~\cite{pope2001turbulent}; see Appendix~\ref{appdx:equivariance} for more details. Physical constraints such as conservation laws can also be included through an augmented loss function - the optimization target during training~\cite[e.g.,][]{raissi2019physics,zhu2019physics,beucler2021enforcing,wu2020enforcing,charalampopoulos2021machine}. Finally, physical constraints can also be enforced in the neural network architecture, e.g., by modifying particular layers~\cite{zanna2020data,mohan2020embedding} or using special components such as equivariance-preserving spatial transformers~\cite{chattopadhyay2020deep,wang2020incorporating,chattopadhyay2021towards}.\\

Building on these earlier studies, here we aim to examine the effectiveness of three methods for incorporating physics into the learning of non-local, data-driven SGS closure models, with a particular focus on the performance in the small-data regime. The three methods employed here are:
\begin{enumerate}
  \item DA, for incorporating rotational equivariances into the input and output training samples,
  \item Physics-constrained loss function that enforces a global enstrophy constraint (EnsCon),
  \item Group CNN (GCNN), a type of equivariance-preserving CNN that has rotational symmetries (equivariances) built into its architecture.
\end{enumerate}
The test case here is a deterministically forced 2D-HIT flow. As discussed in the paper, the use of DA and GCNN are inspired by an example showing the inability of a CNN to account for rotational equivariances in the small-data regime, while the use of EnsCon is motivated by the success of similar global energy constraints in improving the performance of reduced-order and closure models in past studies~\cite{achatz1997closure,charalampopoulos2021machine}. We examine the accuracy of the learnt closure models in \textit{a priori} (offline) tests, in terms of both predicting the SGS terms and capturing inter-scale transfers, and the accuracy and stability of LES with these SGS models in \textit{a posteriori} (online) tests, with regard to long-term statistics. \\

The remaining sections of this paper are as follows. Governing equations of the forced 2D-HIT system, the filtered equations, and the DNS and LES numerical solvers are presented in Section~\ref{sec:eqs}. The CNNs and their {\it a priori} performance trained in the small- and big-data regimes are discussed in Section~\ref{sec:method}. The physics-constrained CNN models (DA, GCNN, and EnsCon) are described in Section~\ref{sec:GCNN}. Results of the \textit{a priori} and \textit{a posteriori} tests with these CNNs in the {\it small} data regime are shown in Section~\ref{sec:results}. Conclusions and future work are discussed in Section~\ref{sec:conclusion}.

\section{DNS and LES: Equations, numerical solvers, and filterings} \label{sec:eqs}
\subsection{Governing equations}
\begin{figure}[t]
\vspace{.1in}
 \centering
 \vspace*{3mm}
 \begin{overpic}[width=0.9\linewidth,height=0.722\linewidth]{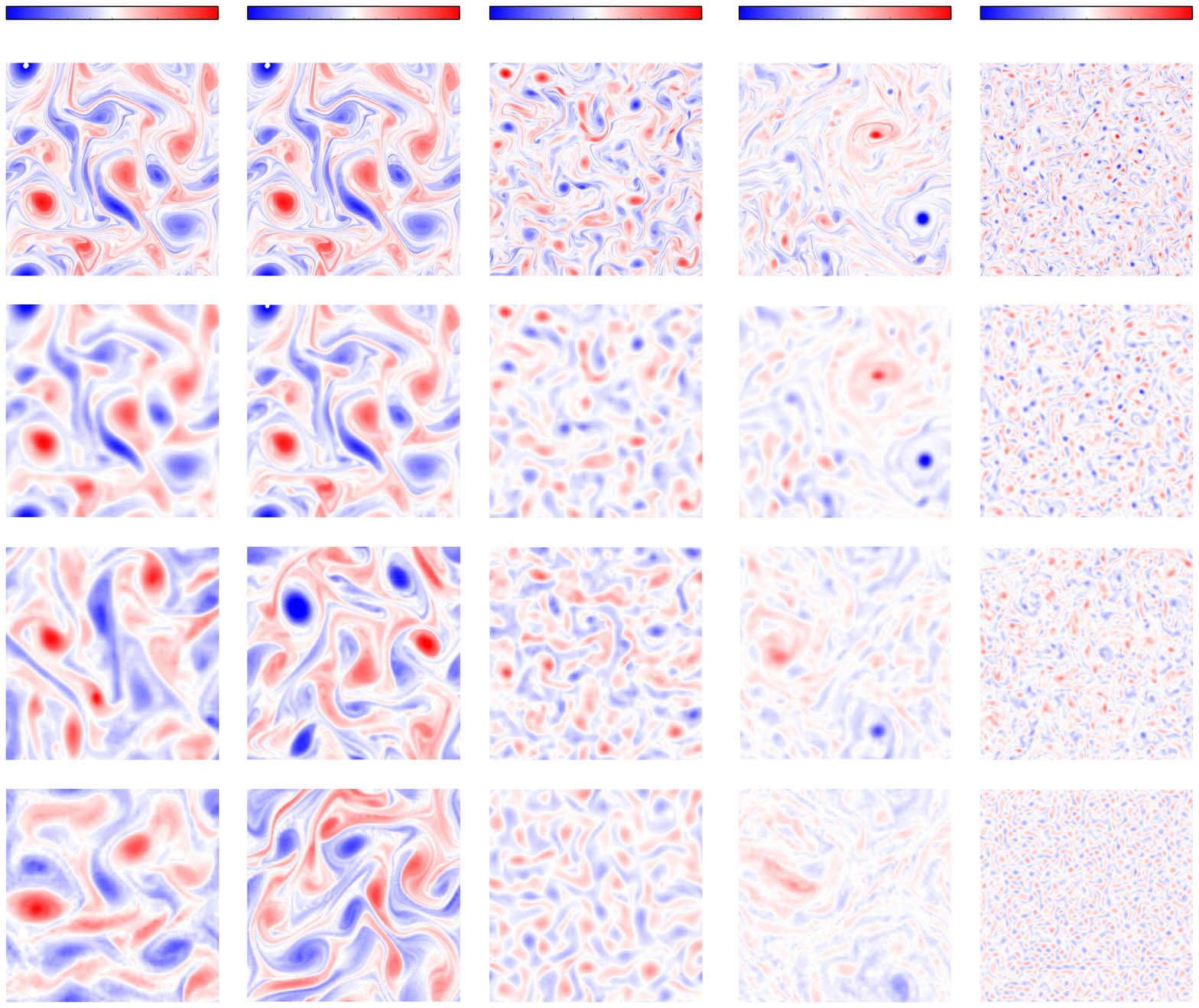}

 \put(-8,9){\scriptsize{DSMAG}}
 \put(-6,28.0){\scriptsize{CNN}}
 \put(-7,47){\scriptsize{FDNS}}
 \put(-6,67){\scriptsize{DNS}}

 \put(0,83.5){\scriptsize{Case $1$: $k_f=4$}}
 \put(0,81){\scriptsize{$r=0.1$, $N_{LES}=64$}}
 \put(0,77.4){\scriptsize{$-25$}}
 \put(8.7,77.4){\scriptsize{$0$}}
 \put(16.3,77.4){\scriptsize{$25$}}

 \put(20,83.5){\scriptsize{Case $2$: $k_f=4$}}
 \put(20,81){\scriptsize{$r=0.1$, $N_{LES}=128$}}
 \put(20.2,77.4){\scriptsize{$-25$}}
 \put(28.95,77.4){\scriptsize{$0$}}
 \put(36.6,77.4){\scriptsize{$25$}}

 \put(40.5,83.5){\scriptsize{Case $3$: $k_f=10$}}
 \put(40.5,81){\scriptsize{$r=0.1$, $N_{LES}=64$}}
 \put(40.5,77.4){\scriptsize{$-50$}}
 \put(49.25,77.4){\scriptsize{$0$}}
 \put(56.8,77.4){\scriptsize{$50$}}

 \put(61.5,83.5){\scriptsize{Case $4$: $k_f=10$}}
 \put(61.5,81){\scriptsize{$r=0.01$, $N_{LES}=64$}}
 \put(61.5,77.4){\scriptsize{$-60$}}
 \put(70.15,77.4){\scriptsize{$0$}}
 \put(77.8,77.4){\scriptsize{$60$}}

 \put(81.8,83.5){\scriptsize{Case $5$: $k_f=25$}}
 \put(81.8,81){\scriptsize{$r=0.1$, $N_{LES}=128$}}
 \put(81.8,77.4){\scriptsize{$-80$}}
 \put(90.48,77.4){\scriptsize{$0$}}
 \put(98,77.4){\scriptsize{$80$}}

 \end{overpic}
  \vspace*{3mm}
 \caption{\small (a) Examples of vorticity fields of DNS, FDNS, and LES with SGS terms modeled by CNN and DSMAG for 5 cases of forced 2D turbulence with different forcing wavenumber $k_f$, friction coefficient $r$, and LES resolution $N_{LES}$. For all cases, $Re=20,000$ and $N_{DNS}=1024$. The scales of the flow structures depend on $k_f$; the higher the $k_f$ the smaller the scales. The linear drag coefficient $r$ determines the similarity of the flow structure. When $r=0.1$, the flow contains several large vortices of similar sizes, while with $r=0.01$ (Case 4), often two large vortices rotating in opposite directions dominate, co-existing with smaller vortices. For the LES, the CNN is trained on big data (number of training snapshots: $n_{tr}=2000, c_{in}<0.75$) to ensure accuracy and numerical stability. DSMAG, in general, captures the large-scale structures but underpredicts the vorticity magnitudes due to the excessive dissipation produced by the non-negative eddy viscosity.}
 \label{fig:1}
\end{figure}

As the testbed, we use forced 2D-HIT, which is a fitting prototype for many large-scale geophysical and environmental flows (where rotation and/or stratification dominate the dynamics). This system has been widely used as a testbed for novel techniques, including ML-based SGS modeling \cite[e.g.,][]{vallis2017atmospheric,chandler2013invariant,thuburn2014cascades,verkley2019maximum,kochkov2021machine}. The dimensionless governing equations in the vorticity ($\omega$) and streamfunction ($\psi$) formulation in a doubly periodic square domain with length $L=2\pi$ are:\\
\begin{subequations}\label{Navier-Stokes}
\begin{eqnarray}
\frac{\partial \omega}{\partial t} + \mathcal{N}(\omega,\psi)&=&\frac{1}{Re}\nabla^2\omega - f -r\omega, \label{eq:NS1}\\
\nabla^2\psi &=& -\omega. \label{eq:NS2}
\end{eqnarray}
\end{subequations}
Here, $\mathcal{N}(\omega,\psi)$ represents the nonlinear advection term:\\
\begin{eqnarray}\label{NL}
\mathcal{N}(\omega,\psi)&=&\frac{\partial \psi}{\partial y}\frac{\partial \omega}{\partial x} - \frac{\partial \psi}{\partial x}\frac{\partial \omega}{\partial y},
\end{eqnarray}
and $f$ represents a deterministic forcing~\cite[e.g.,][]{chandler2013invariant,kochkov2021machine}:
\begin{eqnarray}\label{forcing}
f(x,y)&=& k_f[\cos{(k_fx)} + \cos{(k_fy)}]. \label{eq:forcing}
\end{eqnarray}
We study 5 cases, in which the the forcing wavenumber ($k_f$) and linear friction coefficient ($r$) have been varied, creating a variety of flows that differ in dominant length scales and energy/enstrophy cascade regimes (Figure~\ref{fig:1}). For all cases, the Reynolds number ($Re$) is set to $20,000$. In DNS, as discussed in Section~\ref{sec:numerics}, Eqs.~(\ref{eq:NS1})-(\ref{eq:NS2}) are numerically solved at high spatio-temporal resolutions. \\

To derive the equations for LES, we apply Gaussian filtering~\cite{pope2001turbulent,sagaut2006large,guan2021stable}, denoted by $\overline{(\cdot)}$, to Eqs.~(\ref{eq:NS1})-(\ref{eq:NS2}) to obtain
\begin{subequations}\label{Fitlerd-Navier-Stokes}
\begin{eqnarray}
\frac{\partial \overline{\omega}}{\partial t} + \mathcal{N}(\overline{\omega},\overline{\psi})&=&\frac{1}{Re}\nabla^2\overline{\omega}-\overline{f}-r\overline{\omega}+\underbrace{\mathcal{N}(\overline{\omega},\overline{\psi}) - \overline{\mathcal{N}({\omega},{\psi})}}_{\Pi}\label{eq:FNS1},\\
\nabla^2\overline{\psi} &=& -\overline{\omega}\label{eq:FNS2}.
\end{eqnarray}
\end{subequations}
The LES can be solved using a coarser resolution (compared to DNS) with the SGS term $\Pi$ being the unclosed term, requiring a closure model.\\

\subsection{Numerical simulations}\label{sec:numerics}
In DNS, we solve Eqs.~\eqref{eq:NS1}-\eqref{eq:NS2}. A Fourier-Fourier pseudo-spectral solver is used along with second-order Adams-Bashforth and Crank-Nicolson time-integration schemes for the advection and viscous terms, respectively~\cite{guan2021stable}. The computational grid has uniform spacing $\Delta_\mathrm{DNS} = L/N_\mathrm{DNS}$, where $N_\mathrm{DNS}=1024$ is the number of grid points in each direction. The time-stepping size is set as $\Delta t_\mathrm{DNS}= 5\times10^{-5}$ dimensionaless time unit for all cases except for Case 5, for which $\Delta t_\mathrm{DNS} = 2\times10^{-5}$ is used. For each case, using different random initial conditions, we conducted 3 independent DNS runs to generate the training, offline testing, and online testing datasets. Once the flow reaches statistical equilibrium after a long-term spin-up, each DNS run produces $2000$ snapshots, with each consecutive snapshots $1000\Delta t_\mathrm{DNS}$ apart to reduce the correlation between training samples (inter-sample correlation $c_{in}$, or the correlation coefficient between two consecutive $\Pi$, is below $0.75$; see Section~\ref{sec:method} for further discussions). We have also retained data sampled at $25\Delta t_\mathrm{DNS}$ intervals to examine the effect of $c_{in}$. \\

For LES, we solve Eqs.~\eqref{eq:FNS1}-\eqref{eq:FNS2} employing the same numerical solver used for DNS, but with grid resolutions $N_\mathrm{LES}$ $(=N_\mathrm{DNS}/16 \; \mathrm{or} \; N_\mathrm{DNS}/8)$ listed in Figure~\ref{fig:1}; for each case $\Delta t_\mathrm{LES}= 10\Delta t_\mathrm{DNS}$. The SGS term $\Pi$ is parameterized using a data-driven closure model that is a physics-agnostic or physics-constrained CNN (Sections~\ref{sec:method} and~\ref{sec:GCNN}) or a physics-based dynamic Smagorinsky model (DSMAG). For DSMAG, positive clipping is used to enforce non-negative eddy-viscosity, thus providing stable {\it a posteriori} LES~\cite{pawar2020priori,guan2021stable}.\\

\subsection{Filtered DNS (FDNS) data}\label{sec:FDNS}
To obtain the FDNS and to construct the training and testing data for data-driven methods, we apply a Gaussian filter and then coarse-grain the filtered variables to the LES grid, generating $\overline{\psi}$, $\overline{\omega}$, and $\Pi$~\cite{pope2001turbulent,sagaut2006large}. The filtering and coarse-graining process is described in detail in our recent paper~\cite{guan2021stable}, and is only briefly described here. i) {\it Spectral transformation}: transform the DNS variables into the spectral space by Fourier transform; ii) {\it Filtering}: apply (element-wise-multiply) a Gaussian filter kernel (with filter size $\Delta_F = 2\Delta_\text{LES}$) in the spectral space to {\it filter} the high-wavenumber structures (the resulting variables still have the DNS resolution); iii) {\it Coarse-graining}: truncate the wavenumbers greater than the cut-off wavenumber ($k_c=\pi/\Delta_\text{LES}$) of the filtered variables in the spectral space ((the resulting variables have the LES resolution); iv) {\it Spectral transformation}: transfer the filtered, coarse-grained variables back to the physical space by inverse Fourier transform.\\

\section{Convolutional neural network (CNN): Architecture and results}\label{sec:method}

\subsection{Architecture}
In this work, we first parameterize the unclosed SGS term $\Pi$ in \eqref{eq:FNS1} using a physics-agnostic CNN (CNN hereafter) described in this section. 
The CNN used in this work has the same architecture as the one used in our previous study~\cite{guan2021stable}, which is 10-layer deep with fully convolutional layers, i.e., no pooling or upsampling. All layers are randomly initialized and trainable. The convolutional depth is set to be $64$, and the convolutional filter size is $5\times 5$. We have performed extensive trial and error analysis for these hyperparameters to prevent over-fitting while maintaining accuracy. For example, a CNN with more than 12 layers overfits on this dataset while a CNN with less than 8 layers results in significantly lower {\it a priori} correlation coefficients. The activation function of each layer is the rectified linear unit (ReLU) except for the final layer, which is a linear map. \\

We have standardized the input samples as
\begin{eqnarray}\label{CNN-input}
\Bigg\{\overline{\psi}/\sigma_{ \overline{\psi}},\overline{\omega}/\sigma_{ \overline{\omega}}\Bigg\}\in \mathbb{R}^{2\times N_{\text{LES}}\times N_{\text{LES}}},
\end{eqnarray}
and the output samples as
\begin{eqnarray}\label{CNN-output}
\Bigg\{\Pi/\sigma_{ \Pi}\Bigg\}\in \mathbb{R}^{N_{\text{LES}}\times N_{\text{LES}}},
\end{eqnarray}
where $\sigma_{\overline{\psi}}$, $\sigma_{\overline{\omega}}$, and $\sigma_{\Pi}$ are the standard deviations of $\overline{\psi}$, $\overline{\omega}$, and $\Pi$ calculated over all training samples, respectively. In the later sections, we omit $\sigma$ for clarity, but we always standardize the input/output samples. The CNN is trained as an optimal map $\mathbb{M}$ between the inputs and outputs
\begin{eqnarray}\label{CNN-mapping}
\mathbb{M}:\Big\{\overline{\psi}/\sigma_{\overline{\psi}},\overline{\omega}/\sigma_{\overline{\omega}}\Big\} \in \mathbb{R}^{2\times N_{\text{LES}}\times N_{\text{LES}}} \rightarrow \Big\{\Pi/\sigma_{\Pi}\Big\} \in \mathbb{R}^{N_{\text{LES}}\times N_{\text{LES}}},
\end{eqnarray}
by minimizing the mean-squared-error ($MSE$) loss function
\begin{eqnarray}
MSE = \frac{1}{n_{tr}}\sum_{i=1}^{n_{tr}}\parallel \Pi_i^{\text{CNN}}-\Pi_i^{\text{FDNS}}\parallel_2^2, \label{eq:mse}
\end{eqnarray}
where $n_{tr}$ is the number of training samples and $\parallel \cdot \parallel_2$ is the $L_2$ norm.\\

\subsection{Results}~\label{sec:CNN-results}
\begin{figure}[H]
\vspace{.1in}
 \centering
 \begin{overpic}[width=0.9\linewidth,height=1.2\linewidth]{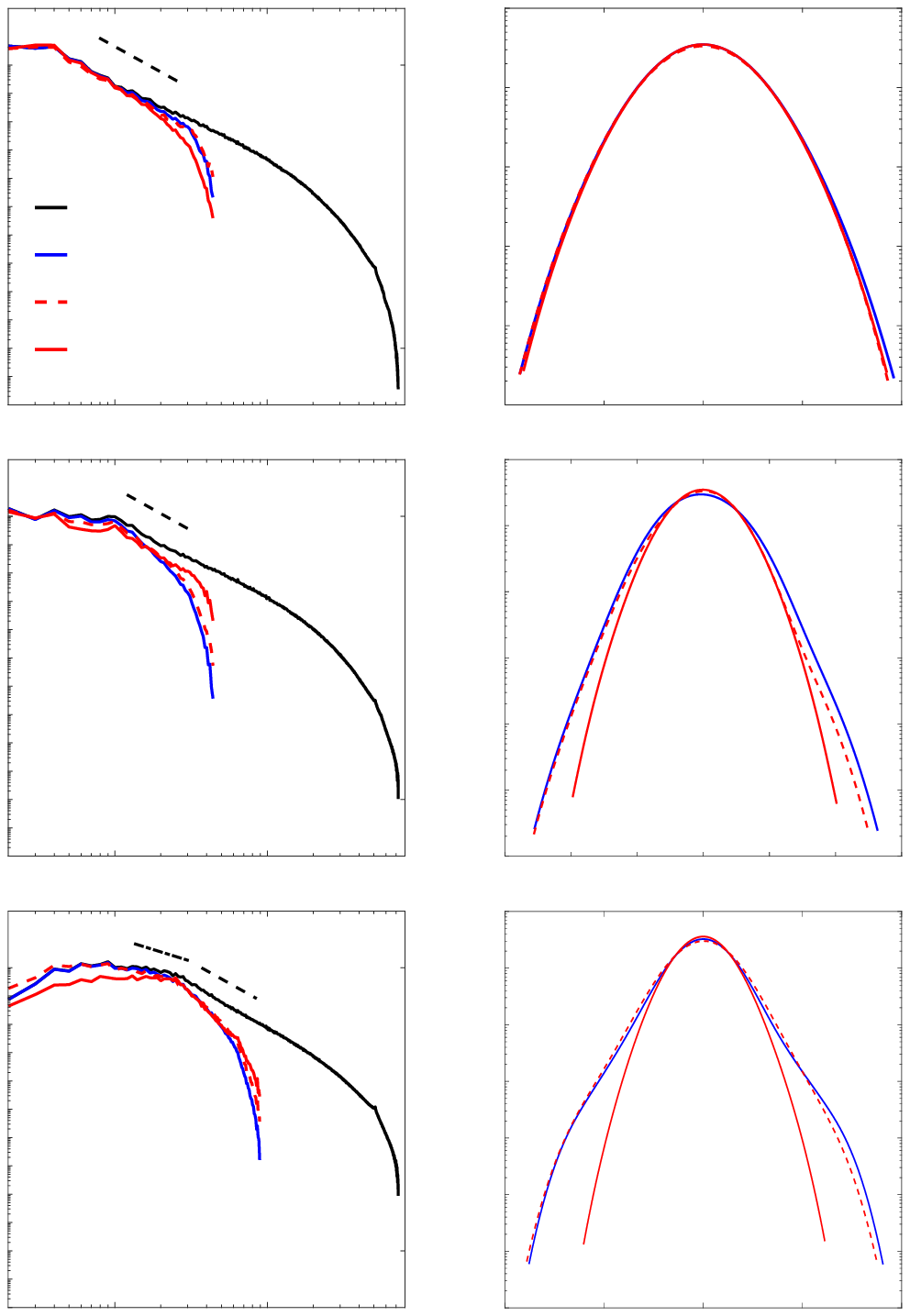}
 \put(12,-4){{$k=(k_x^2+k_y^2)^{0.5}$}}
  \put(56.0,-4){{$\overline{\omega}/\sigma_{\overline{\omega}}$}}
 \put(17,102){\Large{$\hat{E}(k)$}}
 \put(55.5,102){\Large PDF}

 \put(25,95){\large{Case $1$}}
 \put(6,83.8){{DNS}}
 \put(6,80.3){{FDNS}}
 \put(6,76.5){{CNN}}
 \put(6,73){{DSMAG}}
 \put(15,96){{$k^{-3}$}}
 \put(8.5,67){{$10$}}
 \put(21,67){{$10^2$}}
 \put(-3,94.5){{$10^0$}}
 \put(-3.5,84){{$10^{\shortminus5}$}}
 \put(-4,73){{$10^{\shortminus10}$}}

 \put(25,60.5){\large{Case $4$}}
 \put(15,61.5){{$k^{-3}$}}
 \put(8.5,32.5){{$10$}}
 \put(21,32.5){{$10^2$}}
 \put(-3,60){{$10^0$}}
 \put(-3.5,49){{$10^{\shortminus5}$}}
 \put(-4,38.5){{$10^{\shortminus10}$}}

 \put(25,26){\large{Case $5$}}
 \put(20,25.5){{$k^{-3}$}}
 \put(9,28){{$k^{-5/3}$}}
 \put(8.5,-2){{$10$}}
 \put(21,-2){{$10^2$}}
 \put(-3,25.5){{$10^0$}}
 \put(-3.5,15){{$10^{\shortminus5}$}}
 \put(-4,4){{$10^{\shortminus10}$}}

 \put(65,95){\large{Case $1$}}
 \put(38,99){{$10^{\shortminus1}$}}
 \put(38,93){{$10^{\shortminus2}$}}
 \put(38,87){{$10^{\shortminus3}$}}
 \put(38,80.8){{$10^{\shortminus4}$}}
 \put(38,74.8){{$10^{\shortminus5}$}}
 \put(38,68.8){{$10^{\shortminus6}$}}

 \put(41,67.5){{-10}}
 \put(49.0,67.5){{-5}}
 \put(57.8,67.5){{0}}
 \put(66,67.5){{5}}
 \put(73.6,67.5){{10}}

 \put(65,60.5){\large{Case $4$}}
 \put(38,64){{$10^{\shortminus1}$}}
 \put(38,59){{$10^{\shortminus2}$}}
 \put(38,54.3){{$10^{\shortminus3}$}}
 \put(38,49){{$10^{\shortminus4}$}}
 \put(38,44){{$10^{\shortminus5}$}}
 \put(38,39){{$10^{\shortminus6}$}}
 \put(38,34){{$10^{\shortminus7}$}}
 \put(40.5,32.5){{-15}}
 \put(46,32.5){{-10}}
 \put(52,32.5){{-5}}
 \put(57.9,32.5){{0}}
 \put(63.4,32.5){{5}}
 \put(68,32.5){{10}}
 \put(73.5,32.5){{15}}

 \put(65,26){\large{Case $5$}}
 \put(38,29.5){{$10^{\shortminus1}$}}
 \put(38,25.5){{$10^{\shortminus2}$}}
 \put(38,21){{$10^{\shortminus3}$}}
 \put(38,17){{$10^{\shortminus4}$}}
 \put(38,12.5){{$10^{\shortminus5}$}}
 \put(38,8){{$10^{\shortminus6}$}}
 \put(38,4){{$10^{\shortminus7}$}}
 \put(38,0.0){{$10^{\shortminus8}$}}
 \put(41,-1.5){{-10}}
 \put(49,-1.5){{-5}}
 \put(57.8,-1.5){{0}}
 \put(66,-1.5){{5}}
 \put(73.6,-1.5){{10}}
 \end{overpic}
  \vspace*{5mm}
 \caption{\footnotesize Turbulent kinetic energy (TKE, $\hat{E}(k)$) spectra and probability density functions (PDFs) of $\overline{\omega}$ for  representative cases (1, 4, and 5). The CNNs used here are trained on big data. The spectra and PDFs are averaged over 100 samples from the testing set. The PDF is calculated using a kernel estimator~\cite{wilcox2010fundamentals}.}
 \label{fig:2}
\end{figure}

Figure~\ref{fig:1} shows examples of vorticity fields from DNS and FDNS, and from {\it a posteriori} LES that uses CNN or DSMAG for the 5 cases. The CNN used here is trained on the full dataset ($2000$ snapshots with $c_{in}<0.75$), which we will refer to as ``big data'' hereafter. Qualitatively, the LES with CNN more closely reproduces the small-scale features of FDNS compared to DSMAG. To better compare the {\it a posteriori} performance, Figure~\ref{fig:2} shows the turbulent kinetic energy (TKE) spectra $\hat{E}(k)$ and the probability density functions (PDF) of $\overline{\omega}$ averaged over $100$ randomly chosen snapshots from LES (spanning $2\times 10^5\Delta t_{\text{LES}}$ or equally $2\times 10^6\Delta t_{\text{DNS}}$) for 3 representative cases (1, 4, and 5). Note that in forced 2D turbulence, according to the classic Kraichnan-Leith-Batchelor (KLB) similarity theory~\cite{batchelor1969computation,kraichnan1967inertial,leith1968diffusion,perezhogin2019stochastic}, the energy injected by the forcing at wavenumber $k_f$ is transferred to the larger scales ($k<k_f$, energy inverse cascade) while the enstrophy redistributes to the smaller scales ($k>k_f$, enstrophy forward cascade). The KLB theory predicts a $k^{-5/3}$ slope of the TKE spectrum for $k<k_f$ and $k^{-3}$ slope for $k>k_f$.\\

\begin{figure}[t]
\vspace{.1in}
 \centering
 \begin{overpic}[width=0.9\linewidth,height=0.51\linewidth]{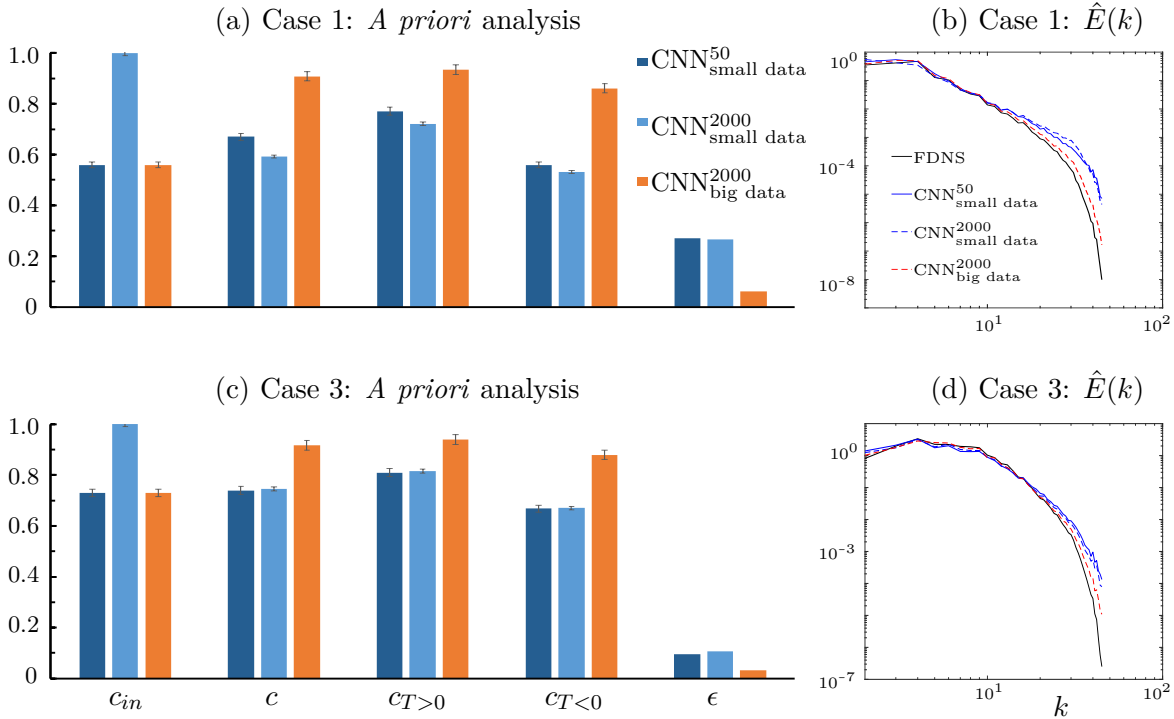}

 \put(15,58.5){(a) Case 1: \emph{A priori} analysis}
 \put(15,25.5){(c) Case 3: \emph{A priori} analysis}
 \put(79,58.5){(b) Case 1: $\hat{E}(k)$}
 \put(79,25.5){(d) Case 3: $\hat{E}(k)$}

 \put(77.5,46.5){\tiny{FDNS}}
 \put(77.5,43.2){\tiny{CNN$_{\text{small data}}^{50}$}}
 \put(77.5,39.8){\tiny{CNN$_{\text{small data}}^{2000}$}}
 \put(77.5,36.5){\tiny{CNN$_{\text{big data}}^{2000}$}}


 \put(54,55){\small{CNN$_{\text{small data}}^{50}$}}
 \put(54,49){\small{CNN$_{\text{small data}}^{2000}$}}
 \put(54,44){\small{CNN$_{\text{big data}}^{2000}$}}

 \put(-2,0){\footnotesize 0}
 \put(-3.5,4.5){\footnotesize 0.2}
 \put(-3.5,9.0){\footnotesize 0.4}
 \put(-3.5,13.4){\footnotesize 0.6}
 \put(-3.5,17.8){\footnotesize 0.8}
 \put(-3.5,22.2){\footnotesize 1.0}

 \put(-2,33){\footnotesize 0}
 \put(-3.5,37.3){\footnotesize 0.2}
 \put(-3.5,42.1){\footnotesize 0.4}
 \put(-3.5,46.5){\footnotesize 0.6}
 \put(-3.5,51.1){\footnotesize 0.8}
 \put(-3.5,55.5){\footnotesize 1.0}

 \put(70,20){\tiny $10^{0}$}
 \put(69,11){\tiny $10^{\shortminus3}$}
 \put(69,-0.5){\tiny $10^{\shortminus7}$}

 \put(70,55.0){\tiny $10^{0}$}
 \put(69,45.5){\tiny $10^{\shortminus4}$}
 \put(69,35){\tiny $10^{\shortminus8}$}

 \put(83,-1.5){\tiny{$10^1$}}
 \put(98,-1.5){\tiny{$10^2$}}

 \put(83,31.5){\tiny{$10^1$}}
 \put(98,31.5){\tiny{$10^2$}}

 \put(5.5,-2){$c_{in}$}
 \put(19.5,-2){$c$}
 \put(30.5,-2){$c_{T>0}$}
 \put(44.5,-2){$c_{T<0}$}
 \put(59,-2){$\epsilon$}
 \put(90,-3){$k$}
 \end{overpic}
  \vspace*{5mm}
 \caption{\small (a) and (c): {\it a priori} and (b) and (d): {\it a posteriori} analyses with 3 CNNs: CNN$_{\text{small data}}^{50}$ uses $n_{tr}=50$ with each two consecutive snapshots being $1000\Delta t_{\text{DNS}}$ apart leading to $c_{in} \sim 0.6-0.7$, CNN$_{\text{small data}}^{2000}$ uses $n_{tr}=2000$ with each two consecutive snapshots being $25\Delta t_{\text{DNS}}$ apart, leading to $c_{in} \sim 0.99$, and CNN$_{\text{big data}}^{2000}$ uses $n_{tr}=2000$ with each two consecutive snapshots being $1000\Delta t_{\text{DNS}}$ apart, leading to $c_{in} \sim 0.6-0.7$. The training DNS datasets of CNN$_{\text{small data}}^{50}$ and CNN$_{\text{small data}}^{2000}$ span the same time range ($50,000\Delta t_{\text{DNS}}$), and their performance are very similar both in \emph{a priori} and \emph{a posteriori} analyses. The training DNS datasets of CNN$_{\text{big data}}^{2000}$ are 40 times longer (``big data''), and these models outperform the ones trained on ``small data'' based on all the {\it a priori} metrics in (a) and (c) and {\it a posteriori} performance in (b) and (d). The error bars denote plus and minus standard deviation (the error bars on $\epsilon$ are small and not shown for the save of clarify).}
 \label{fig:3}
\end{figure}

In general, the $\hat{E}(k)$ of LES with CNN better matches the FDNS than that of the LES with DSMAG. For Cases 1 and 4, where the enstrophy forward cascade dominates, the LES with DSMAG incorrectly captures the spectra at small scales. For Case 5, where the energy inverse cascade is important too, the DSMAG fails to recover the energy at large scales correctly. Examining the PDFs of $\overline{\omega}$ shows that in Cases 4 and 5, the PDF from LES with CNN almost overlaps with the one from FDNS even at the tails, while the PDF from LES with DSMAG deviates beyond $3$ standard deviations. Due to the excessive dissipation, the LES with DSMAG is incapable of capturing the extremes (tails of the PDF). Therefore, in {\it a posteriori} analysis, similar to decaying 2D turbulence~\cite{guan2021stable}, for different setups of forced 2D turbulent flows, LES with CNN {\it trained with big data} better reproduces the FDNS flow statistics as compared to LES with DSMAG. Note that in this study, we are comparing the CNN-based closures against DSMAG, which is more accurate and powerful than the typical baseline, the static Smagorinsky model~\cite{maulik2019subgrid,guan2021stable}. Finally, we highlight that the CNN has outstanding {\it a priori} performance too, yielding $c>0.9$ (Figure~\ref{fig:3}). \\

Although the CNNs yield outstanding performance in both {\it a priori} and {\it a posteriori} analyses when trained with {\it big data}, their performance  deteriorates when the training dataset is {\it small}. Before introducing three physics-constrained CNNs for overcoming this problem (Section~\ref{sec:GCNN}), we first show in Figure~\ref{fig:3} classifying that ``big" versus ``small" data depends  not only on the number of snapshots in the training dataset ($n_{tr}$) but also on the inter-sample correlation ($c_{in}$). In \emph{a priori} analysis (bar plots in Figure~\ref{fig:3}), we use four metrics. The first two are $c_{in}$, which is the average correlation coefficient between consecutive snapshots of $\Pi$ in the training set, and $c$, which is the average correlation coefficient between the true (FDNS) and CNN-predicted $\Pi$ over $100$ random snapshots in the testing set. Following past studies~\cite{maulik2019subgrid,guan2021stable}, we introduce
\begin{eqnarray}\label{SGStransfer}
T = sgn(\nabla^{2}\overline\omega)\odot \Pi,
\end{eqnarray}
whose sign at a grid point determines whether the SGS term is diffusive $T>0$ or anti-diffusive $T<0$ ($\odot$ denotes element-wise multiplication). The third metric we use is $c$ computed separately based on the sign of $T$ for testing samples: $c_{T>0}$, which is the average $c$ on the grid points experiencing diffusion by SGS processes and $c_{T<0}$, which is the average $c$ on the grid points experiencing anti-diffusion. Finally, noticing that the global enstrophy transfer due to the SGS term is $\langle\overline{\omega}\Pi\rangle$ (see Appendix~\ref{appdx}), where $\langle \cdot \rangle$ denotes domain averaging, we define $\epsilon$, the relative error in global enstrophy transfer by the SGS term as
\begin{eqnarray}
\epsilon = |\langle\overline{\omega}\Pi^{\text{CNN}}\rangle-\langle\overline{\omega}\Pi^{\text{FDNS}}\rangle|/|\langle\overline{\omega}\Pi^{\text{FDNS}}\rangle|.\label{eq:eps}
\end{eqnarray}

Figure~\ref{fig:3} compares the performance of CNNs trained with three training sets for two representative cases (1 and 3). CNN$_{\text{small data}}^{50}$ uses $n_{tr}=50$ with each two consecutive snapshots being $1000\Delta t_{\text{DNS}}$ apart, leading to $c_{in} \sim 0.6-0.7$, while CNN$_{\text{small data}}^{2000}$ uses $n_{tr}=2000$ with each two consecutive snapshots being $25\Delta t_{\text{DNS}}$ apart, leading to highly correlated samples with $c_{in}\approx 0.99$ (note that these two DNS datasets have the same total time length). CNN$_{\text{big data}}^{2000}$ uses $n_{tr}=2000$ with each two consecutive snapshots being $1000\Delta t_{\text{DNS}}$ apart, leading to $c_{in} \sim 0.6-0.7$ (note that this datasets is 40 times longer than the other two). \\

The {\it a priori} results show that CNN$_{\text{small data}}^{50}$ and CNN$_{\text{small data}}^{2000}$ have comparable $c$, $c_{T>0}$, $c_{T<0}$, and $\epsilon$, which are all worse than those of CNN$_{\text{big data}}^{2000}$. This demonstrates the importance of both $n_{tr}$ and $c_{in}$ in determining the effective size of the training set and the performance of the learnt closure (note that similar to what was shown in our earlier work~\cite{guan2021stable}, in the small-data regime, $c_{T<0}<c_{T>0}$). The {\it a posteriori} analysis leads to the same conclusion: The TKE spectra of LES with CNN$_{\text{big data}}^{2000}$ closely matches that of the FDNS while the spectra of LES with CNN$_{\text{small data}}^{50}$ and CNN$_{\text{small data}}^{2000}$ are comparable and do not match the spectra of FDNS at wavenumbers larger than around $10$.\\

These analyses show that increasing the number of training samples from $50$ to $2000$ within the same DNS dataset does not enhance the performance of CNN. In general, the performance of CNN depends on the total DNS time length that the training dataset spans. In fact, both the number of training snapshots ($n_{tr}$) and the inter-sample correlation ($c_{in}$) determine whether we are in the ``big" or ``small" data regime. As ``big" datasets may not be available for many problems, in the next section, we will discuss how to enhance the performance of the CNNs in the small-data regime using physics constraints.\\


\pagebreak

\section{Physics-constraint CNNs: Incorporating rotational equivariance and SGS enstrophy transfer}\label{sec:GCNN}
In this section, we demonstrate how incorporating rotational equivariances via DA or GCNNs, or enforcing a global SGS enstrophy transfer in the loss function can improve the {\it a priori} and {\it a posteriori} performance of the CNN-based closures in the {\it small-data} regime (with $n_{tr}=50)$. \\

\begin{figure}
\vspace{.1in}
 \centering
 \begin{overpic}[width=0.9\linewidth,height=0.35\linewidth]{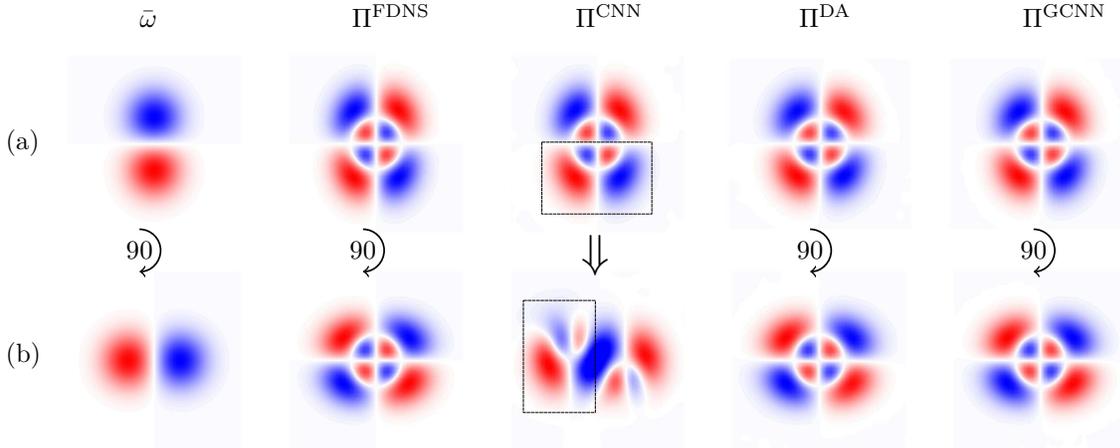}
 \put(-3,29){\small{(a)}}
 \put(-3,10){\small{(b)}}
 \put(9,40){\small{$\bar{\omega}$}}
 \put(28,40){\small{$\Pi^{\text{FDNS}} $}}
 \put(48,40){\small{$\Pi^{\text{CNN}}$}}
 \put(68,40){\small{$\Pi^{\text{DA}} $}}
 \put(88,40){\small{$\Pi^{\text{GCNN}}$}}

 \put(7,18){\small{$\CricArrowRight{90}$}}
 \put(27,18){\small{$\CricArrowRight{90}$}}
 \put(48.5,18.5){\LARGE{$\Downarrow$}}
 \put(67,18){\small{$\CricArrowRight{90}$}}
 \put(87,18){\small{$\CricArrowRight{90}$}}

 \end{overpic}
 \caption{\small A dipole vortex shows the shortcoming of a physics-agnostic CNN in capturing the rotational equivariance of the SGS term (third column). The physics-agnostic CNN regards the rotational transform between the training and testing vortex field as a translational transform (the translation of the structure in the black dashed box). However, the CNN with DA or GCNN can capture this rotational equivariance correctly (fourth and fifth columns). The symbol \protect\scalebox{0.45}{$\protect\CricArrowRight{90}$} means rotation by 90$^{\circ}$ clockwise and $\Downarrow$ means translation.}
 \label{fig:4}
\end{figure}

The example in Figure~\ref{fig:4} shows one of the shortcomings of a physics-agnostic CNN: the inability to capture rotational equivariance when the training set is small. In this example, a CNN is trained on snapshots of $\bar{\omega}$ and its $\Pi$ term (from Gaussian filtering) for an inviscid vertically aligned vortex dipole (first and second columns from left, first row). This dipole moves around the domain, i.e., it translates but does not rotate (thus, a horizontally aligned dipole is never seen in the training set). The trained CNN can accurately predict the $\Pi$ term for out-of-sample $\bar{\omega}$ snapshots (third column for left, first row). However, for a $\bar{\omega}$ snapshot that is rotated by 90$^{\circ}$ clockwise (first column, second row), the CNN cannot accurately predict the correct $\Pi$ term, which is also rotated by 90$^{\circ}$ clockwise (second column, second row). Note that there was no horizontally aligned dipole like this in the training set. The CNN, instead, predicts an incorrect $\Pi$ term that is based on separate translations of the two parts of the vertically aligned dipole (third column, second row)\footnote{The CNN basically predicts that because the red blob of the dipole is now to the left of the blue blob, the part of the $\Pi$ term corresponding to the red blob in the first row now should be to the right of the part corresponding to the blue blob (see the black box in the third column).}.\\

The implication of this example is that if the training set only involves limited flow configurations (i.e., small-data regime) such as only those from the first row, then the CNN can be quite inaccurate for a testing set involving new configurations such as those in the second row. In the much more complicated 2D-HIT flow, there are various complex flow configurations. In a big training set, it is more likely that these different configurations would be present and the CNN learns their corresponding $\Pi$ terms and the associated transformations; however, this is less likely in a small training set. 
The SGS term $\Pi$ is known to be equivariant to translation and rotation, i.e., if the flow state variables are translated or rotated, $\Pi$ should also be translated or rotated to the same degree~\cite{pope2001turbulent}. While translation equivariance is already achieved in a regular CNN by weight sharing~\cite{cohen2016group}, rotational equivariance is not guaranteed. Recent studies show that rotational equivariance can actually be critical in data-driven SGS modeling~\cite{prat2020priori,frezat2021physical,prakash2021invariant,pawar2022frame}. To capture the rotational equivariance in the small-data regime, we propose two separate approaches: (1) DA, by including 3 additional rotated (by 90$^{\circ}$, 180$^{\circ}$, and 270$^{\circ}$) counterparts of each original FDNS snapshot in the training set~\cite{frezat2021physical} and (2) by using a GCNN architecture, which enforces rotational equivariance by construction~\cite{cohen2016group,bronstein2021geometric}.\\


 The GCNN uses group convolutions, which increases the degree of weight sharing by transforming and reorienting the filters such that the feature maps in GCNN are equivariant under imposed symmetry transformations, e.g., rotation and reflection~\cite{veeling2018rotation}. In our work, the group convolutional filters are oriented at 0$^{\circ}$, 90$^{\circ}$, 180$^{\circ}$, and 270$^{\circ}$ such that the feature map and the output ($\Pi^{\text{GCNN}}$) are rotationally equivariant with respect to the inputs ($\bar{\psi}$ and $\bar{\omega}$), i.e., Eq.~\eqref{eq:CNN-equivariance} in Appendix~\ref{appdx:equivariance}. \\

In addition to the structural modeling approaches mentioned above that achieve rotational equivariance (still with the MSE loss function, Eq.~\eqref{eq:mse}), we can also modify the loss function to combine structural and functional modeling approaches to enhance the performance of CNN in the small-data regime. For example, in 2D turbulence, the SGS enstrophy transfer is critical in maintaining the accuracy and stability of LES~\cite{perezhogin20202d,maulik2019subgrid,guan2021stable}. Therefore, capturing the correct SGS enstrophy transfer $\langle\overline{\omega}\Pi\rangle$ in a CNN can be important for its performance. The {\it a priori} analysis in Figure~\ref{fig:3} already showed that the error in capturing the global SGS enstrophy transfer $\epsilon$ is small in the big-data regime, but can be large in the small-data regime. Here we propose to add a penalty term to the loss function that acts as regularization, enforcing (as a soft constraint) the global SGS enstrophy transfer. Similar to the global energy constraint implemented in Ref.~\cite{charalampopoulos2021machine}, this physics-constrained loss function consist of the MSE plus the global SGS enstrophy transfer error:\\
\begin{eqnarray}
Loss = \frac{1-\beta}{n_{tr}}\sum_{i=1}^{n_{tr}}\parallel \Pi_i^{\text{CNN}}-\Pi_i^{\text{FDNS}}\parallel_2^2 +  \frac{\beta}{n_{tr}}\sum_{i=1}^{n_{tr}}|\langle\overline{\omega}_i\Pi_i^{\text{CNN}}\rangle-\langle\overline{\omega}_i\Pi_i^{\text{FDNS}}\rangle|,\label{eq:loss}
\end{eqnarray}
where $\beta\in[0,1]$ is an adjustable hyperparameter. We empirically find $\beta=0.5$ to be optimal in minimizing the relative total enstrophy transfer error ($\epsilon$) without significantly affecting $c$. This  physics-constrained loss function (Eq.~\eqref{eq:loss})  synergically combines the structural and functional modeling approaches.\\

\section{Results}\label{sec:results}
\subsection{\emph{A priori} analysis}\label{sec:a-priori}
\begin{figure}[t]
\vspace{.1in}
 \centering
 \begin{overpic}[width=0.9\linewidth,height=0.5\linewidth]{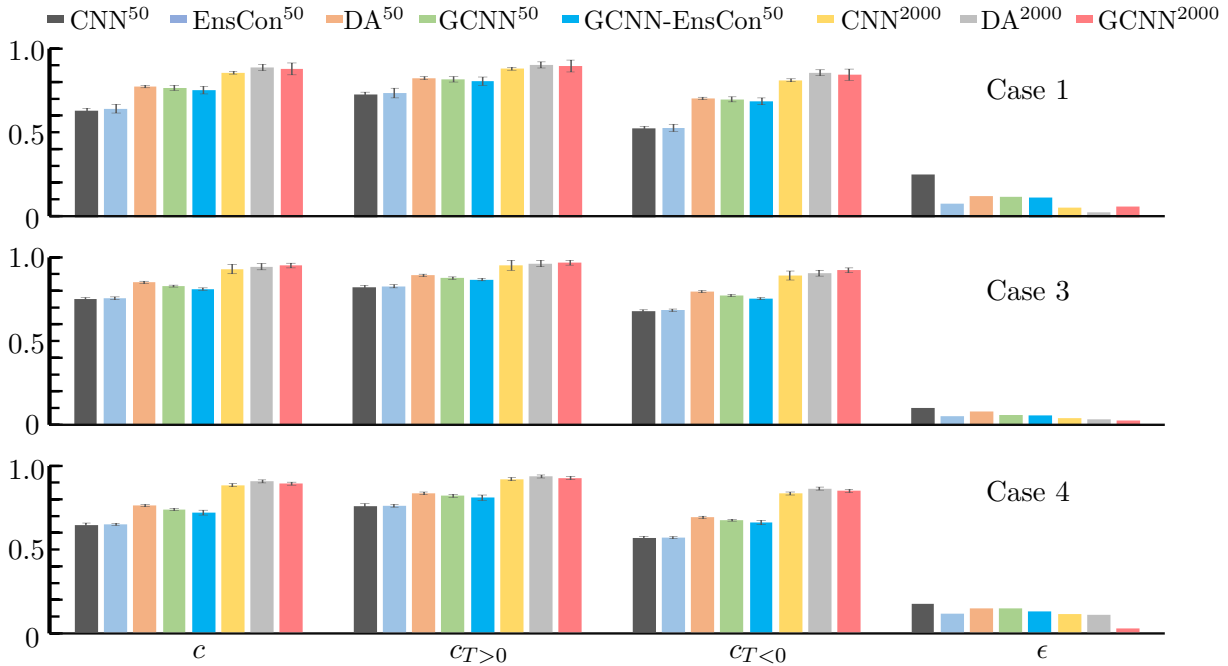}

 \put(84,48){Case 1}
 \put(84,30){Case 3}
 \put(84,12){Case 4}

 \put(2,54.3){\small{CNN$^{50}$}}
 \put(13,54.3){\small{EnsCon$^{50}$}}
 \put(26.5,54.3){\small{DA$^{50}$}}
 \put(35,54.3){\small{GCNN$^{50}$}}
 \put(48,54.3){\small{GCNN-EnsCon$^{50}$}}
 \put(71,54.3){\small{CNN$^{2000}$}}
 \put(83.5,54.3){\small{DA$^{2000}$}}
 \put(94,54.3){\small{GCNN$^{2000}$}}

 \put(-2,-1){0}
 \put(-3.5,6.8){0.5}
 \put(-3.5,14){1.0}

\put(-2,18){0}
\put(-3.5,25.3){0.5}
\put(-3.5,33){1.0}

\put(-2,36.5){0}
\put(-3.5,44){0.5}
\put(-3.5,51.5){1.0}


 \put(13,-2){$c$}
 \put(36,-2){$c_{T>0}$}
 \put(61,-2){$c_{T<0}$}
  \put(88.5,-2){$\epsilon$}

 \end{overpic}
  \vspace*{5mm}
 \caption{\small \emph{A priori} analysis in terms of correlation coefficients $c$, $c_{T>0}$, and $c_{T>0}$, and the relative enstrophy transfer error $\epsilon$. It is shown that EnsCon does not improve the structural modeling metric $c$ but significantly reduces the functional modeling error, $\epsilon$. DA and GCNN enhance the structural modeling performance and also reduce $\epsilon$. The superscripted number denotes the $n_{tr}$ in the training dataset with $c_{in}<0.75$ as in Figure~\ref{fig:3}. The error bars denote plus and minus one standard deviation (the error bars on $\epsilon$ are small and not shown for the sake of clarify). }
 \label{fig:5}
\end{figure}

{\it A priori} analysis is performed using the following metrics: correlation coefficients ($c$), global enstrophy transfer error ($\epsilon$), and scale-dependent enstrophy and energy transfers ($T_Z$ and $T_E$, as defined later in this section).  Figure~\ref{fig:5} shows the bar plots of $c$, $c_{T<0}$, $c_{T>0}$, and $\epsilon$ for 3 representative cases (1, 3, and 4). In the small-data regime ($n_{tr}=50$), the use of DA or GCNN increases the correlation coefficients $c$, $c_{T<0}$, and $c_{T>0}$; the increases are largest for $c_{T>0}$, whose low values could lead to instabilities in {\it a posteriori} LES, as discussed earlier. The use of DA or GCNN also decreases the relative total enstrophy transfer error $\epsilon$, particularly for Case~1. One point to highlight here is that DA can achieve the same, and in some cases even better, {\it a priori} accuracy compared to GCNN, while the network architecture is much simpler in DA, which builds equivariance simply in the training data. The EnsCon does not improve the correlation coefficient $c$ (because it only adds a functional modeling component), but as expected, it decreases $\epsilon$, which as shown later improves the \emph{a posteriori} LES. To examine a combined approach, we build an enstrophy-constrained GCNN (GCNN-EnsCon), which performs somewhere in between GCNN and EnsCon: GCNN-EnsCon has a higher $c$ than EnsCon but lower than GCNN, and GCNN-EnsCon has higher $\epsilon$ than EnsCon but lower than GCNN. As shown later, the LES with GCNN-EnsCon has the best {\it a posteriori} performance among all tested models. \\

To summarize Figure~\ref{fig:5}, the physics-constrained CNNs trained on small data ($n_{tr}=50$) outperform the physics-agnostic CNN trained on small data, but none could outperform the physics-agnostic CNN trained on 40 times more data ($n_{tr}=2000$) in these {\it a priori} tests. However, we emphasize that $40$ is a substantial factor in terms of the amount of high-fidelity data. This figure also shows that adding physics constraints to the CNN trained in the big-data regime ($n_{tr}=2000$) does not necessarily lead to any improvement over CNN$^{2000}$, suggesting that these physics constraints could be learnt by a physics-agnostic CNN from the data given enough training samples. \\

To further assess the performance of the closures computed using physics-constrained CNNs trained in the small-data regime, we also examine the scale-dependent enstrophy and energy transfers ($T_Z$ and $T_E$) defined as:~\cite{thuburn2014cascades,perezhogin2021priori}\\
 \begin{eqnarray}
 T_Z(k) = \mathfrak{R}(-\hat{\Pi}^*_{k}\hat{\bar{\omega}}_{k}),\label{eq:TZ}\\
 T_E(k) = \mathfrak{R}(-\hat{\Pi}^*_{k}\hat{\bar{\psi}}_{k}). \label{eq:TE}
\end{eqnarray}
Here, $\mathfrak{R}(\cdot)$ means the ``real part of", $\hat{(\cdot)}$ denotes Fourier transform, and the asterisks denote complex conjugate. The scale-dependent enstrophy/energy transfer is positive for enstrophy/energy backscatter (enstrophy/energy moving from subgrid scales to resolved scales) and negative for enstrophy/energy forward transfer (enstrophy/energy moving from resolved scales to subgrid scales)~\cite{perezhogin2021priori}. Note that backscatter and forward transfer here are inter-scale transfers by the SGS term ($\Pi$), and are different concepts from inverse and forward cascades (discussed earlier in Section~\ref{sec:CNN-results}).\\

\begin{figure}[t]
\vspace{.1in}
 \centering
 \begin{overpic}[width=0.8\linewidth,height=0.48\linewidth]{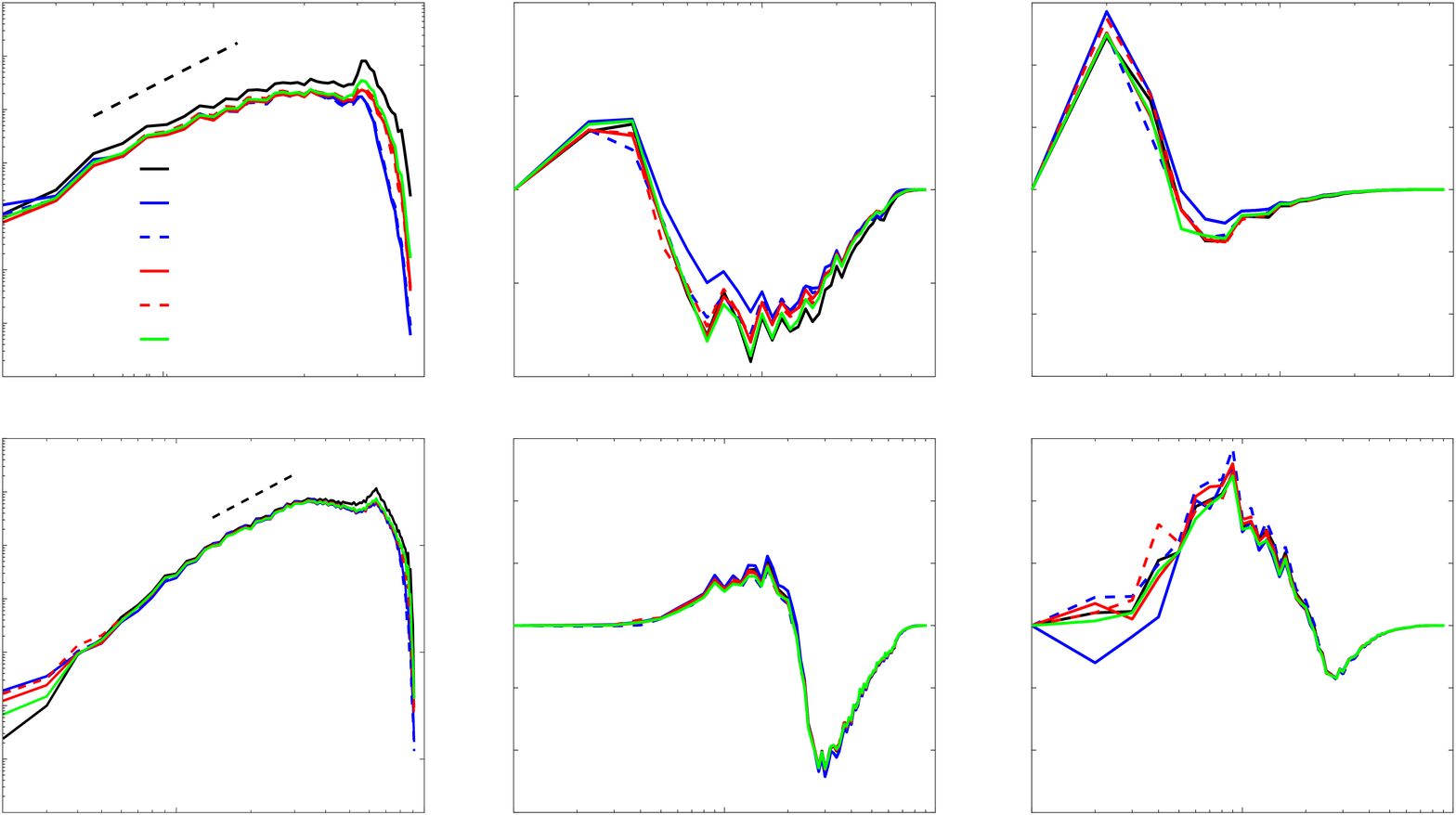}
\put(-10,25){\small{Case 5}}
\put(-10,58){\small{Case 1}}

\put(24,57.5){\small{(a)}}
\put(59,57.5){\small{(b)}}
\put(95,57.5){\small{(c)}}
\put(24,24.9){\small{(d)}}
\put(59,24.9){\small{(e)}}
\put(95,24.9){\small{(f)}}

 \put(11,62){$|\hat{\Pi}(k)|^2$}
 \put(47,62){$T_Z(k)$}
 \put(82,62){$T_E(k)$}
 \put(13,-5){$k$}
 \put(51,-5){$k$}
 \put(82,-5){$k$}

 \put(11,22.5){\small{$k^{2.46}$}}
 \put(8,55){\small{$k^{2.46}$}}
 \put(12,47){\tiny{FDNS}}
 \put(12,44.5){\tiny{CNN$^{50}$}}
 \put(12,42){\tiny{EnsCon$^{50}$}}
 \put(12,39.5){\tiny{DA$^{50}$}}
 \put(12,37){\tiny{GCNN$^{50}$}}
 \put(12,34.5){\tiny{GCNN-EnsCon$^{50}$}}

 \put(-3,19){\tiny{$10^{0}$}}
 \put(-4,11){\tiny{$10^{\shortminus2}$}}
 \put(-4,3){\tiny{$10^{\shortminus4}$}}
 \put(11,-2){\tiny{$10$}}
 \put(27,-2){\tiny{$100$}}

 \put(34,22.5){\tiny{$2$}}
 \put(34,18){\tiny{$1$}}
 \put(34,13.5){\tiny{$0$}}
 \put(33,8.5){\tiny{$\shortminus1$}}
 \put(33,4){\tiny{$\shortminus2$}}
 \put(48,-2){\tiny{$10$}}
 \put(62,-2){\tiny{$100$}}

 \put(68,22.5){\tiny{$0.4$}}
 \put(68,18){\tiny{$0.2$}}
 \put(69,13.5){\tiny{$0$}}
 \put(67,8.5){\tiny{$\shortminus0.2$}}
 \put(67,4){\tiny{$\shortminus0.4$}}
 \put(84,-2){\tiny{$10$}}
 \put(98,-2){\tiny{$100$}}

 \put(-3,51){\tiny{$10^{0}$}}
 \put(-4,43){\tiny{$10^{\shortminus2}$}}
 \put(-4,35){\tiny{$10^{\shortminus4}$}}
 \put(10,30.5){\tiny{$10$}}
 \put(28,30.5){\tiny{$50$}}

 \put(32.5,52.5){\tiny{$0.1$}}
 \put(34,45.5){\tiny{$0$}}
 \put(31.5,38.5){\tiny{$\shortminus0.1$}}
 \put(51,30.5){\tiny{$10$}}
 \put(63,30.5){\tiny{$50$}}

 \put(67,54.5){\tiny{$10^{\shortminus2}$}}
 \put(69,45.5){\tiny{$0$}}
 \put(66,36){\tiny{$\shortminus10^{\shortminus2}$}}
 \put(87,30.5){\tiny{$10$}}
 \put(99,30.5){\tiny{$50$}}
 \end{overpic}
  \vspace*{5mm}
 \caption{\small \emph{A priori} analysis in terms of scale-dependent power spectra $|\hat{\Pi}(k)|^2$, and scale-dependent enstrophy and energy transfers $T_Z$ and $T_E$ for two representative cases (1 and 5). The inertial part of the $|\hat{\Pi}(k)|^2$ spectrum has a slope of 2.46, consistent with previous studies~\cite{perezhogin2021priori, berner2009spectral}. (a)-(c): In Case 1 where the enstrophy direct cascade is important (as discussed in Section~\ref{sec:CNN-results}), CNN$^{50}$ does not capture the power spectra correctly at high wavenumbers, and the enstrophy forward transfer ($T_Z<0$) is under-predicted. (d)-(f): In Case 5 where the energy inverse cascade is important (see Section~\ref{sec:CNN-results}), the prediction discrepancy occurs at the low wavenumbers of the power spectra, and at the backscattering part of the energy transfer ($T_E>0$). In general, the proposed physics-constrained CNNs (DA, GCNN, EnsCon, and GCNN-EnsCon) reduce the prediction error in both structural ($|\hat{\Pi}(k)|^2$) and functional ($T_Z$ and $T_E$) modelings metrics.}
 \label{fig:6}
\end{figure}

Figure~\ref{fig:6} shows the power spectra of $|\hat{\Pi}(k)|^2$, $T_Z$, and $T_E$ from FDNS and different CNNs, providing further evidence that the incorporated physics constraints improves the {\it a priori} performance of the data-driven closures. For Case 1 (first row), the $|\hat{\Pi}(k)|^2$ is better predicted by DA and GCNN at the high wavenumbers. The scale-dependent enstrophy forward transfer ($T_Z<0$ in Figure~\ref{fig:6}(b)) is underpredicted by CNN, and the deviation from FDNS is corrected by DA, GCNN, EnsCon, and GCNN-EnsCon. For Case 5 (second row), however, where the inverse energy cascade is important (see Section~\ref{sec:CNN-results}), the gain from the physicss-constrained CNNs (DA, GCNN, EnsCon, and GCNN-EnsCon) can be seen in the scale-dependent energy transfer ($T_E$ in Figure~\ref{fig:6}(f)), where the physics-agnostic CNN incorrectly predicts a portion ($k<4$) of energy backscatter ($T_E>0$) to be forward energy transfer ($T_E<0$).

\subsection{\emph{A posteriori} analysis}

\begin{figure}[t]
\vspace{.4in}
 \centering
 \begin{overpic}[width=0.9\linewidth,height=0.4\linewidth]{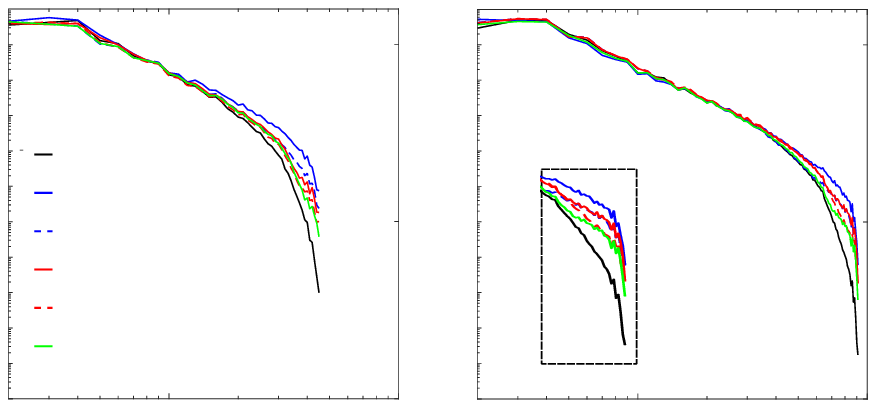}

 \put(35,40){Case 1}
 \put(90,40){Case 2}

 \put(20,46){{$\hat{E}(k)$}}
 \put(25,-4){{$k$}}

 \put(75,46){{$\hat{E}(k)$}}
 \put(80,-4){{$k$}}

 \put(7,27){\footnotesize{FDNS}}
 \put(7,23){\footnotesize{CNN$^{50}$}}
 \put(7,18.5){\footnotesize{EnsCon$^{50}$}}
 \put(7,14){\footnotesize{DA$^{50}$}}
 \put(7,10){\footnotesize{GCNN$^{50}$}}
 \put(7,6){\footnotesize{GCNN-EnsCon$^{50}$}}

 \put(-5,0){\footnotesize{$10^{\shortminus10}$}}
 \put(-3.8,19.5){\footnotesize{$10^{\shortminus5}$}}
 \put(-3,39.5){\footnotesize{$10^{0}$}}
 \put(-1,-2){\footnotesize{$10^{0}$}}
 \put(18,-2){\footnotesize{$10^{1}$}}
 \put(44,-2){\footnotesize{$10^{2}$}}

 \put(49,0){\footnotesize{$10^{\shortminus10}$}}
 \put(50.2,19.5){\footnotesize{$10^{\shortminus5}$}}
 \put(51,39.5){\footnotesize{$10^{0}$}}
 \put(53,-2){\footnotesize{$10^{0}$}}
 \put(72,-2){\footnotesize{$10^{1}$}}
 \put(98,-2){\footnotesize{$10^{2}$}}

 \end{overpic}
  \vspace*{5mm}
 \caption{\small The TKE spectra $\hat{E}(k)$ of Cases 1 and 2 from \textit{a posteriori} LES run. Results are from long-term LES integrations ($2\times 10^5\Delta t_{\text{LES}}$ or $2\times 10^6\Delta t_{\text{DNS}}$). The $\hat{E}(k)$ is calculated from $100$ randomly chosen snapshots and then averaged. The insert in Case 2 magnifies the tails of the $\hat{E}(k)$ spectra for better visualization. In general, the physics-constrained CNNs (DA$^{50}$, GCNN$^{50}$, EnsCon$^{50}$, or GCNN-EnsCon$^{50}$) improve the \textit{a posteriori} performance of LES compared to the LES with the physics-agnostic CNN$^{50}$: The spectra from the LES with physics-constrained CNNs better match the FDNS spectra especially at the tails (high-wavenumber structures). In particular, the spectra from the LES with GCNN-EnsCon$^{50}$ have the best match with the FDNS spectra especially at the high wavenumbers. The improvement is more prominent for the coarser-grid LES (Case 1, $N_{LES}=64$, compared to Case 2, $N_{LES}=128$).}
   \vspace*{5mm}
 \label{fig:7}
\end{figure}

\begin{figure}[t]
\vspace{.1in}
 \centering
 \begin{overpic}[width=0.9\linewidth,height=0.4\linewidth]{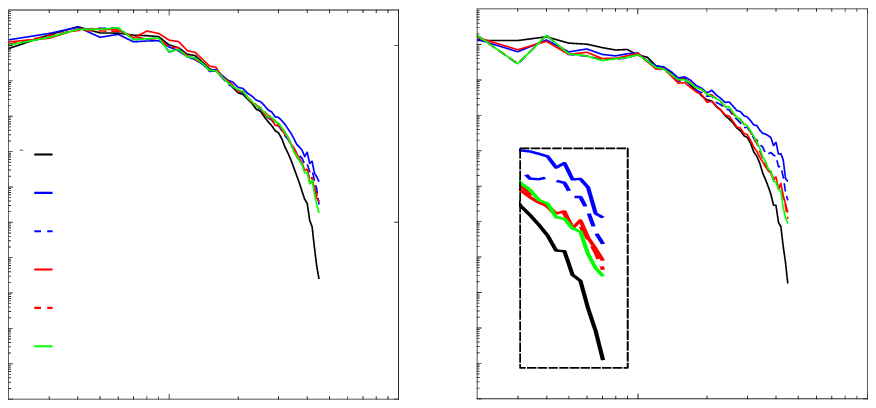}

 \put(35,40){Case 3}
 \put(90,40){Case 4}

 \put(20,46){{$\hat{E}(k)$}}
 \put(25,-4){{$k$}}

 \put(75,46){{$\hat{E}(k)$}}
 \put(80,-4){{$k$}}

 \put(7,27){\footnotesize{FDNS}}
 \put(7,23){\footnotesize{CNN$^{50}$}}
 \put(7,18.5){\footnotesize{EnsCon$^{50}$}}
 \put(7,14){\footnotesize{DA$^{50}$}}
 \put(7,10){\footnotesize{GCNN$^{50}$}}
 \put(7,6){\footnotesize{GCNN-EnsCon$^{50}$}}

 \put(-5,0){\footnotesize{$10^{\shortminus10}$}}
 \put(-3.8,19.5){\footnotesize{$10^{\shortminus5}$}}
 \put(-3,39.5){\footnotesize{$10^{0}$}}
 \put(-1,-2){\footnotesize{$10^{0}$}}
 \put(18,-2){\footnotesize{$10^{1}$}}
 \put(44,-2){\footnotesize{$10^{2}$}}

 \put(49,0){\footnotesize{$10^{\shortminus10}$}}
 \put(50.2,19.5){\footnotesize{$10^{\shortminus5}$}}
 \put(51,39.5){\footnotesize{$10^{0}$}}
 \put(53,-2){\footnotesize{$10^{0}$}}
 \put(72,-2){\footnotesize{$10^{1}$}}
 \put(98,-2){\footnotesize{$10^{2}$}}

 \end{overpic}
  \vspace*{5mm}
 \caption{\small Same as Figure~\ref{fig:7} but for Cases~3 and 4. Similar to the finding of Figure~\ref{fig:7}, the physics-constrained CNNs (DA$^{50}$, GCNN$^{50}$, EnsCon$^{50}$, or GCNN-EnsCon$^{50}$) improve the \textit{a posteriori} performance of LES compared to the LES with the physics-agnostic CNN$^{50}$. For Case~3, the improvement can only be observed at the highest wavenumber. For Case~4, however, incorporating the rotational equivariance (DA$^{50}$, GCNN$^{50}$, and GCNN-EnsCon$^{50}$) leads to a more accurate LES than the enstrophy constraint alone (EnsCon$^{50}$) alone. Also similar to the finding of Figure~\ref{fig:7}, the spectra from the LES with GCNN-EnsCon$^{50}$ have the best match with the FDNS spectra especially at the high wavenumbers.}
 \label{fig:8}
\end{figure}

\begin{figure}[t]
\vspace{.4in}
 \centering
 \begin{overpic}[width=0.9\linewidth,height=0.4\linewidth]{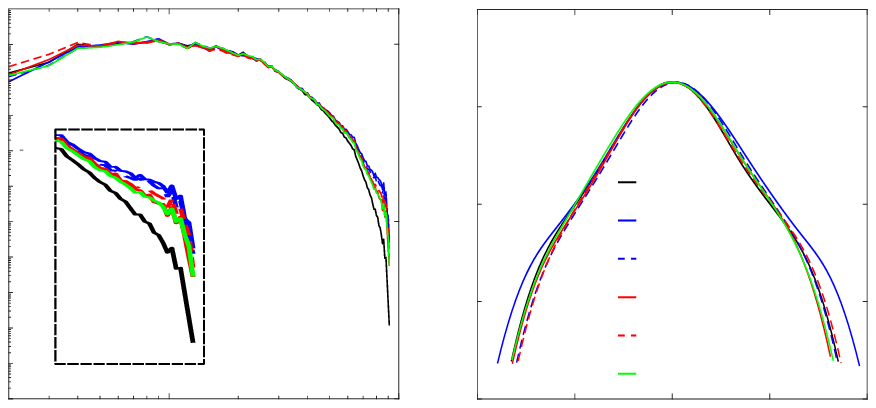}

 \put(35,40){Case 5}
 \put(90,40){Case 5}

 \put(18,46){{(a) $\hat{E}(k)$}}
 \put(25,-4){{$k$}}

 \put(74,46){(b) PDF}
 \put(80,-4){{$\bar{\omega}/\sigma_{\bar{\omega}}$}}

 \put(73.5,24){\footnotesize{FDNS}}
 \put(73.5,20){\footnotesize{CNN$^{50}$}}
 \put(73.5,15.5){\footnotesize{EnsCon$^{50}$}}
 \put(73.5,11){\footnotesize{DA$^{50}$}}
 \put(73.5,6.5){\footnotesize{GCNN$^{50}$}}
 \put(73.5,2.5){\footnotesize{GCNN-EnsCon$^{50}$}}

 \put(-5,0){\footnotesize{$10^{\shortminus10}$}}
 \put(-3.8,19.5){\footnotesize{$10^{\shortminus5}$}}
 \put(-3,39.5){\footnotesize{$10^{0}$}}
 \put(-1,-2){\footnotesize{$10^{0}$}}
 \put(18,-2){\footnotesize{$10^{1}$}}
 \put(44,-2){\footnotesize{$10^{2}$}}

 \put(50.2,0){\footnotesize{$10^{\shortminus8}$}}
 \put(50.2,10){\footnotesize{$10^{\shortminus6}$}}
 \put(50.2,21.5){\footnotesize{$10^{\shortminus4}$}}
 \put(50.2,32){\footnotesize{$10^{\shortminus2}$}}
 \put(51,43){\footnotesize{$10^{0}$}}
 \put(53,-2){\footnotesize{$\shortminus10$}}
 \put(65,-2){\footnotesize{$\shortminus5$}}
 \put(76.6,-2){\footnotesize{$0$}}
 \put(88,-2){\footnotesize{$5$}}
 \put(98,-2){\footnotesize{$10$}}
 \end{overpic}
  \vspace*{5mm}
 \caption{\small Same as Figure~\ref{fig:7} but showing (a) The TKE spectra $\hat{E}(k)$ and (b) probability density function (PDF) of $\overline{\omega}$ for Case 5. Although, the physics-constrained CNNs (DA$^{50}$, GCNN$^{50}$, EnsCon$^{50}$, or GCNN-EnsCon$^{50}$) result in slightly improved LES in terms of the TKE spectrum, the gain from the physical constraints can be observed more clearly in the PDF of vorticity where the LES with CNN$^{50}$ over-predicts the extreme values (tails of the PDF).}
 \label{fig:9}
\end{figure}

Figures~\ref{fig:7}-\ref{fig:9} show the $\hat{E}(k)$ spectra of Cases 1-5 . In general, the TKE spectrum from LES with the physics-agnostic CNN (denoted by CNN$^{50}$) matches the one from FDNS at low wavenumbers (large-scale structures) but severely over-predicts the TKE at high wavenumbers (small-scale structures). For example, CNN$^{50}$ starts to deviate from FDNS at $k\approx 20$ for Cases 1, 3, and 4 as shown in Figure~\ref{fig:7} (left) and Figure~\ref{fig:8}. This over-prediction can lead to unphysical and unstable numerical results. For example, the vorticity field of LES with CNN$^{50}$ exhibits extensive noisy (i.e., very high-wavenumber) structures in several simulations (not shown). All LES runs with physics-constrained CNNs (DA$^{50}$, GCNN$^{50}$, EnsCon$^{50}$, and GCNN-EnsCon$^{50}$) outperform the LES with CNN$^{50}$. In particular, for Cases 2 (Figure~\ref{fig:7} (right)) and 3 (Figure~\ref{fig:8} (left)), the LES runs with DA$^{50}$, GCNN$^{50}$, EnsCon$^{50}$, and GCNN-EnsCon$^{50}$ produce similar TKE spectra which are consistently better than that of the LES with CNN$^{50}$. For Cases 1 (Figure~\ref{fig:7} (left)) and 4 (Figure~\ref{fig:8} (right)), however, incorporating rotational equivariance (through DA, GCNN, or GCNN-EnsCon) leads to better {\it a posteriori} performances than incorporating the global enstrophy constraint alone (EnsCon) in terms of matching the FDNS spectra. Overall, the LES with GCNN-EnsCon$^{50}$ has the best performance in these 4 cases, showing the advantage of combining different types of physics constraints in the small-data regime.\\

In Case 5, the gain from the physics constraints is less obvious from the TKE spectra, although a slight improvement at the tails can still be observed (Figure~\ref{fig:9} (left)). In this case, the PDF of vorticity (Figure~\ref{fig:9} (right)) better reveals the gain, where LES with CNN$^{50}$ predicts spuriously large vorticity extremes due to the excessive high-wavenumber structures in the vorticity field. The physics-constrained CNNs (DA$^{50}$, GCNN$^{50}$, EnsCon$^{50}$, and GCNN-EnsCon$^{50}$) result in a stable and more accurate LES as the TKE spectrum and PDF of vorticity better match those of the FDNS. \\

\section{Summary and discussion}\label{sec:conclusion}
The objective of this paper is to learn CNN-based non-local SGS closures from filtered DNS data for stable and accurate LES, with a focus on the {\it small-data regime}, i.e., when the available DNS training set is small. We demonstrate that incorporating physics constraints into the CNN using three methods can substantially improve the \emph{a priori} (offline) and \emph{a posteriori} (online) performance of the data-driven closure model in the small-data regime.  \\

We use 5 different forced 2D homogeneous isotropic turbulence (HIT) flows with various forcing wavenumbers, linear drag coefficients, and LES grid sizes as the testbeds. First, we show in Section~\ref{sec:method} that in the ``big-data" regime (with $n_{tr}=2000$ weakly correlated training samples), the LES with physics-agnostic CNN is stable and accurate, and outperforms the LES with the physics-based DSMAG closure, particularly as the data-driven closure captures backscattering well (see Figure~\ref{fig:2} and Ref.~\cite{guan2021stable}). Next, we show, using \emph{a priori} (offline) and \emph{a posteriori} (online) tests, that the performance of the physics-agnostic CNNs substantially deteriorate when they are trained in the ``small-data" regime: with $n_{tr}=2000$ highly correlated samples or with $n_{tr}=50$ weakly correlated samples. This analysis demonstrates that the small versus big data regime depends not only on the number of training samples but also on their inter correlations. \\



To improve the performance of CNNs trained in the small-data regime, in Section~\ref{sec:GCNN}, we propose incorporating physics in the CNNs through using 1) data augmentation (DA), 2) a group equivariant CNN (GCNN), or an enstrophy-constrained loss function (EnsCon). The idea behind using DA and GCNN is to account for the rotational equivariance of the SGS term. This is inspired by a simple example of a vortex dipole, which shows that for never-seen-before samples, the physics-agnostic CNN can only capture the translational equivariance, but not the rotational equivariance, another important property of the SGS term. The idea behind EnsCon is to combine structural and functional modeling approaches through a regularized loss function. {\it A priori} and {\it a posteriori} tests show that all these physics-constrained CNNs outperform the physics-agnostic CNN in the small-data regime ($n_{tr}=50$). Improvements of the data-driven SGS closures using the GCNN, which uses an equivariant-preserving architecture, are consistent with the findings of a recent study \citep{pawar2022frame}, though it should be mentioned that DA, which simply builds equivariance in the training samples and can be used with any architecture, shows comparable or even in some cases better performance than GCNN.

Overall, GCNN+EnsCon, which combines these two main constraints, demonstrate the best {\it a posteriori} performance, showing the advantage of adding physics constraints together. Note that here we focus on rotational equivariance, which is a property of the 2D turbulence test case. In other flows, other equivariance properties might exist (e.g., reflection equivariance as in Couette flow and Rayleigh B\'{e}nard convection), and they can be incorporated through DA or GCNN as needed. These results show the major advantage and potential of physics-constrained deep learning methods for SGS modeling in the small-data regime, which is of substantial importance for complex and high-Reynolds number flows, for which the availability of high-fidelity (e.g., DNS) data could be severely  limited.

\section*{Acknowledgments}
This work was supported by an award from the ONR Young Investigator Program (N00014-20-1-2722), a grant from the NSF CSSI program (OAC-2005123), and by the generosity of Eric and Wendy Schmidt by recommendation of the Schmidt Futures program. Computational resources were provided by NSF XSEDE (allocation ATM170020) and NCAR$'$s CISL (allocation URIC0004). Our codes and data are available at~\url{https://github.com/envfluids/2D-DDP}.

\appendix
\section{Appendix: Equivariance properties of the SGS term}
According to the transformation properties of the Navier-Stokes equations~\cite{pope2001turbulent}, the SGS term $\Pi$ should satisfy:
\label{appdx:equivariance}
\begin{eqnarray}\label{eq:equivariance}
\Pi(T_g\omega,T_g\psi)&=&T_g\Pi(\omega,\psi),
\end{eqnarray}
where $T_g$ represents a translational or rotational transformation. $\omega$ and $\psi$ are the vorticity and streamfunction, respectively (as described in Section~\ref{sec:eqs}).\\

In the ML literature, ``equivariance" means that transforming an input (e.g., by translation or rotation, denoted by $T_g$) and then passing the transformed input through the learnt map (CNN in our case) should give the same result as first mapping the input and then transforming the output~\cite{cohen2016group,veeling2018rotation}:
\begin{eqnarray}\label{eq:CNN-equivariance}
\Pi^{\text{CNN}}(T_g\bar{\omega},T_g\bar{\psi},\theta)&=&T_g\Pi^{\text{CNN}}(\bar{\omega},\bar{\psi},\theta).
\end{eqnarray}
Here, $\theta$ represents a group of learnable parameters of the network. To preserve the translational and rotational equivariance of $\Pi$ (Eq.~\eqref{eq:equivariance}), the network parameters $\theta$ should be learnt such that  Eq.~\eqref{eq:CNN-equivariance} is satisfied. In turbulence modeling, ``equivariance" may also be referred to as ``symmetry"~\cite{pope2001turbulent,silvis2017physical}. In this paper, we use the term ``equivariance", and use an equivarience-preserving network (GCNN) or build this property into the training via DA.

\section{Appendix: The global enstrophy-transfer constraint}\label{appdx}
The equation for enstrophy transfer can be obtained by first multiplying the filtered equation~(Eq.~\eqref{eq:FNS1}) by $\overline{\omega}$:
\begin{eqnarray}\label{eq:Enstrophy}
\overline{\omega}\frac{\partial \overline{\omega}}{\partial t} + \overline{\omega}\mathcal{N}(\overline{\omega},\overline{\psi})&=&\overline{\omega}\frac{1}{Re}\nabla^2\overline{\omega}-\overline{\omega}\overline{f}-r\overline{\omega}^2+\underbrace{\overline{\omega}\mathcal{N}(\overline{\omega},\overline{\psi}) - \overline{\omega}\overline{\mathcal{N}({\omega},{\psi})}}_{\overline{\omega}\Pi}.
\end{eqnarray}
Rearranging Eq.~\eqref{eq:Enstrophy} gives:
\begin{eqnarray}\label{eq:Enstrophy2}
\frac{1}{2}\frac{\partial \overline{\omega}^2}{\partial t} + \frac{1}{2}\mathcal{N}(\overline{\omega}^2,\overline{\psi})&=&\frac{1}{Re}\left[\frac{1}{2}\nabla^2\overline{\omega}^2-(\nabla\overline{\omega})^2\right]-\overline{\omega}\overline{f}-r\overline{\omega}^2+\overline{\omega}\Pi.
\end{eqnarray}
The evolution equation for domain-averaged enstrophy $Z=\langle\frac{1}{2}\overline{\omega}^2\rangle$ is then obtained by domain averaging Eq.~\eqref{eq:Enstrophy2} and invoking the domain's periodicity~\cite{davidson2015turbulence}:\\
\begin{eqnarray}\label{eq:Enstrophy3}
\frac{dZ}{dt}&=&-\frac{1}{Re}\langle(\nabla\overline{\omega})^2\rangle-\langle\overline{\omega}\overline{f}\rangle-2rZ+\langle\overline{\omega}\Pi\rangle.
\end{eqnarray}
Therefore, the domain-averaged enstrophy transfer due to the SGS term is $\langle\overline{\omega}\Pi\rangle$. In Eq.~\eqref{eq:loss}, we enforce $\langle\overline{\omega}\Pi\rangle$ predicted by the CNN to be close to that of the FDNS as a domain-averaged (global) soft constraint.

\pagebreak
\setlength{\bibsep}{2.6pt plus 1ex}
\begin{spacing}{.01}
	\small
\bibliographystyle{elsarticle-num-names}
\bibliography{main}
\end{spacing}
\end{document}